\documentclass[a4paper,11pt]{article}
\pdfoutput=1
\usepackage{graphicx}

\usepackage{amsmath}
\usepackage{amssymb}
\usepackage{latexsym}
\usepackage{color}
\usepackage{float}
\usepackage{cite}
\usepackage{makeidx}

\restylefloat{figure}

\newcommand{\bra}{\begin{array}}
\newcommand{\era}{\end{array}}
\newcommand{\beq}{\begin{equation}}
\newcommand{\eeq}{\end{equation}}
\newcommand{\bqr}{\begin{eqnarray}}
\newcommand{\eqr}{\end{eqnarray}}

\def\BC{\bb C}
\def\_\BC{\bbi C}

\def\( {\left(}
   \def\) {\right)}
\def\[ {\left[}
\def\] {\right]}
\def\no2 {{\textstyle{n\over 2}}}

\newcommand{\si}{\sigma}

\newcommand{\lb}{\label}

\begin{document}
\begin{titlepage}
\setcounter{page}{1}
\renewcommand{\thefootnote}{\fnsymbol{footnote}}

\begin{flushright}
\end{flushright}

\vspace{5mm}
\begin{center}

{\Large \bf {
Transport Properties for Triangular Barriers \\ in  Graphene Nanoribbon}}

\vspace{5mm}

{\bf Abderrahim El Mouhafid}$^{a}$ and {\bf Ahmed
Jellal}$^{a,b,c}$\footnote{ajellal@ictp.it, a.jellal@ucd.ac.ma}

\vspace{5mm}

{$^{a}$\em Theoretical Physics Group,  
Faculty of Sciences, Choua\"ib Doukkali University},\\
{\em 24000 El Jadida, Morocco}

{$^{b}$\em Saudi Center for Theoretical Physics, Dhahran, Saudi
Arabia}


{$^c$\em Physics Department, College of Sciences, King Faisal University,\\
Alahssa 31982, Saudi Arabia}


\vspace{3cm}

\begin{abstract}
We theoretically study the electronic transport properties of
Dirac fermions through one and double triangular barriers in
graphene  {nanoribbon}. Using the
transfer matrix method,
we determine the transmission, conductance and Fano factor.
They are obtained to be
various parameters dependent such as well width, barrier
height and barrier width.
Therefore, different discussions are given and
comparison with the previous significant works is done.
In particular,
it is
shown that
at Dirac point the Dirac fermions {always} own a minimum
conductance associated with a maximum Fano factor
and change  their behaviors
in an oscillatory way (irregularly periodical tunneling peaks) when
the potential of applied voltage is increased.

\end{abstract}
\end{center}
\vspace{3cm}

\noindent PACS numbers: 72.80.Vp, 73.21.-b, 71.10.Pm, 03.65.Pm

\noindent Keywords: graphene, scattering, triangular potential,
transmission.
\end{titlepage}


\section{Introduction}
Graphene
 is a single two-dimensional array of carbon atoms with a honeycomb lattice, which was discovered in 2004~\cite{Novoselov}.
This finding has been attracted an intensive attention from both
experimental and theoretical
aspects.
In particular, the tunneling of Dirac fermions in graphene has
already been verified experimentally~\cite{Stander}, which in turn
has spurred an extraordinary amount of interest in the
investigation of the electronic transport properties in graphene
based quantum wells, barriers, p–n junctions, transistors, quantum
dots, superlattices, etc. The electrostatic barriers in graphene
can be generated in various ways~\cite{Katsnelsonn, Sevinçli}. For
example, it can be done by applying a gate voltage, cutting it
into finite width nanoribbons and using doping
or otherwise. Whereas 
magnetic barrier {can, in principle, 
can be realized by using magnetic strips or using superconductors \cite{group1}.}
As far  graphene, results of the transmission
coefficient and the tunneling conductance were already reported
for the electrostatic
barriers
\cite{Sevinçli, Anna, Biswas, Tworzydle, Alhaidari, Novoselovvv}
and magnetic barriers~\cite{Masir, Choubabi, Jellal}.

The electronic band structure (energy dispersion relation) of
graphene consists of two inequivalent pairs of cones with apices
located at Brillouin-zone corners. The dispersion relation $E=\pm
\hbar v_{F}|\vec{k}|$ is linear around the Dirac point ($K$, $K'$)
where {$v_F\simeq 10^{6}$m/s} is Fermi
velocity~\cite{Neto}. The presence of such Dirac-like
quasiparticles is expected to induce some unusual electronic
properties, which make difference with respect to 
two-dimensional electronic gas, such as the so-called Klein
paradox~\cite{Klein}, anomalous integer quantum Hall
effect~\cite{Novoselovv, Gusynin, Nomura} and observation of minimum
conductivity~\cite{Gusynin}. The fact that in an ideal graphene
sheet the carriers are massless, gives rise to Klein paradox,
which allows particles to tunnel through any electrostatic
potential barriers, that is the wavefunction has an oscillatory
tail outside the electrostatic barrier region. Hence this property
excludes the possibility to confine electrons using electrostatic
gates, as in usual semiconductors. Thus to enable the fabrication
of confined structures, such as quantum dots, we need to use
another type of {barrier  such as the infinite mass
barrier~\cite{Berry}.}

Theoretical investigations have been widely performed to clarify
the resonant-tunneling features using 
mostly 
barriers of the rectangular forms. The reasons because the
corresponding models are so simple to have an advantage for
numerical calculations. 
However few works studied tunneling effect 
 with barriers of the  potential slopes as a
result of externally applied field~\cite{Brennan, Allen,
mouhafid, Alhaidari2}. One of them is the trapezoidal double barrier
structure, which was investigated to study the effect of the
potential disturbance at the interfaces of the graphene cheet~\cite{Inaba}.
In the same spirit,
we consider another problem based on 
single and double
triangular barrier structures.
Our model
is possibly applied to the
resonant tunneling diodes of which the barriers are formed by
delta doping in the future and is also a step outward from a
rectangular form from the other point of view.
We ensure the confinement of Dirac fermions in the $y$-direction
by using infinite mass confinement, which requires infinite mass
at the boundary of the $y$-strip and results in a specific
quantization of the $y$-component of the momentum~\cite{Berry}.
The effects of the
well width,  barrier height and  barrier width on the transport properties
are
systematically studied through numerical calculations.
As long as the applied potential is increased, 
the number of the minimum
conductance associated with  maximum Fano factor increases as well. This
result makes difference with respect to that
of rectangular barrier where there is only one minimum and one maximum
\cite{Tworzydle}.
We conclude that it is relatively more easily for
Dirac fermions to tunnel through a triangular barrier in a
graphene sheet rather than rectangular one.

The outline of the paper is the following. In section 2, we set our
theoretical model by giving the appropriate equation  describing
Dirac fermions in graphene and choosing the convenient
configuration for 
the triangle double barrier structures as depicted in Figure
\ref{fig01.eps}. In section 3, we expose the exact analytical
solution to solve the Dirac equation in each
regions of the 
structures, which resulted in giving the corresponding eigenvalues
and eigenspinors. Tunneling probabilities are calculated in
section 4 as a functions of different parameters such as the
fermion energy, static electric field and  incident angle. These
are done by matching spinors in different interfaces and
using the transfer matrix techniques.
In section 5, we discuss  the transport results corresponding to
single and double barriers separately. The obtained
results show characteristic oscillations associated with tunneling
resonances as a function of the fermion energy and the static
electric field. We conclude our work in the final section.

\section{Theoretical formulation}

\begin{figure}[h]
\begin{center}
\includegraphics[width=2.5in]{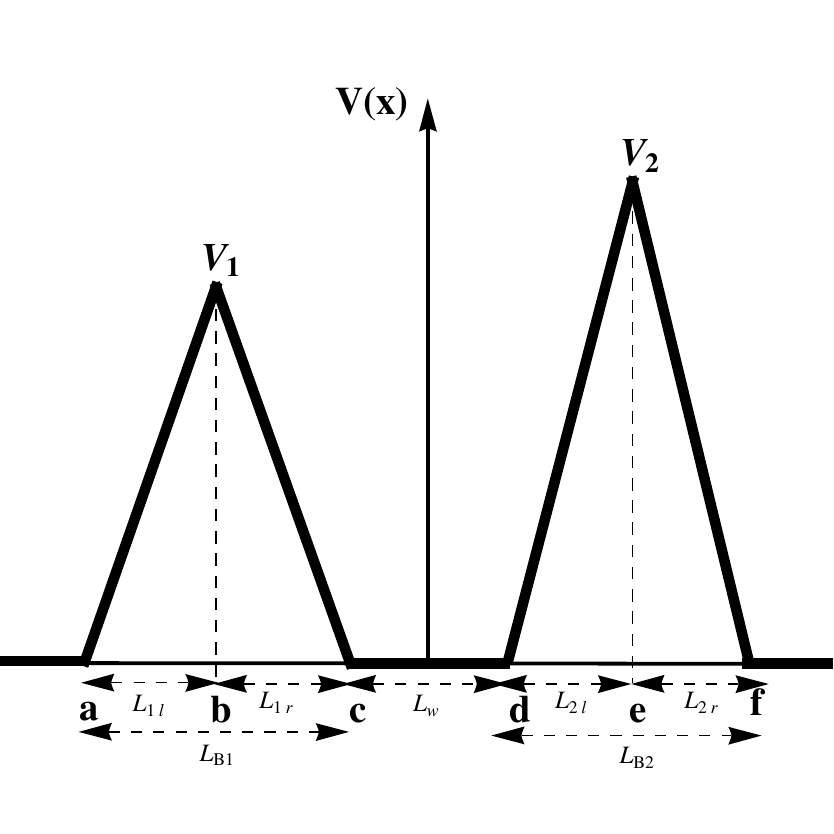}
\caption{\sf The parameters of a triangle double barrier
structure.} \label{fig01.eps}
\end{center}
\end{figure}

We consider a system of  massless
Dirac fermions through a strip of graphene characterized by the
length $L_B$ and width $w$ in the presence of a double triangular
barriers. In the systems made of graphene, the two Fermi points,
each with a two-fold band degeneracy, can be described by a
low-energy continuum approximation with a four-component envelope
wavefunction whose components are labeled by a Fermi-point
pseudospin $=\pm 1$ and a sublattice forming an honeycomb. Being a
zero-gap semiconductor, the quasiparticle motion can be described
by the {massless} Dirac like equation 
\begin{eqnarray}\label{ak}
[v_F \vec{\sigma}\cdot\vec{p}+V(x)]\psi(x,y)=\varepsilon\psi(x,y)
\end{eqnarray}
where $v_F\simeq9.84\times10^{6}m/s$ is the Fermi velocity,
$\vec{p}=-i\hbar\overrightarrow{\nabla}$ is the momentum operator
(hereafter $v_F=\hbar=1$), $\vec{\sigma}=(\sigma_{x},\sigma_{y})$
are the Pauli matrices, $\varepsilon=\hbar v_F|\vec{k}|$ being the
energy of the incident particle. A triangular double barrier
configuration is depicted in Figure \ref{fig01.eps} with all
parameters, which requires two kinds of width: the right and left
sides of the barrier. Therefore the dependence of the various
parameters can be considered as shown in the
potential $V(x)$ configuration 
\begin{eqnarray}\label{ak}
 V(x)=\left\{
        \begin{array}{ll}
          (x-a)F_{1}, & \qquad x\in [a,b] \\
          (x-c)F_{2}, & \qquad x\in [b,c] \\
          (x-d)F_{3}, & \qquad x\in [d,e] \\
          (x-f)F_{4}, &  \qquad x\in [e,f] \\
          0, & \qquad {\mbox{otherwise}}
        \end{array}
      \right.
\end{eqnarray}
where we have set $F_{1}=\frac{V_1}{b-a}$, $F_{2}=\frac{V_1}{b-c}$,
$F_{3}=\frac{V_2}{e-d}$ and $F_{4}=\frac{V_2}{e-f}$ are the
strength of the static electric field in each regions. 

Our system is supposed to have finite width $w$ with infinite mass
boundary conditions on the wavefunction at the boundaries $y = 0$
and $y=w$ along the $y$-direction 
\cite{Alhaidari, Novoselovvv, Masir, Choubabi, Jellal, Berry}.
This boundary conditions result in a quantization of the
transverse momentum along the $y$-direction, which is
\begin{equation}\lb{1}
k_{y}=k_{n}=\frac{\pi}{w}\left(n+\frac{1}{2}\right), \qquad
n=0,1,2\cdots .
\end{equation}
One can therefore assume a spinor solution of the following form
$\psi_{j}(x,y)=(\phi_{j_{+}}(x),\phi_{j_{-}}(x))^{t}e^{ik_{y}y}$
where $j=1$ for $x<a$, $2$ for $x\in[a,b]$, $3$ for $x\in[b,c]$,
$4$ for $x\in[c,d]$, $5$ for $x\in[d,e]$, $6$ for $x\in[e,f]$ and
$7$ for $x>f$ denotes the different space regions. Thus our
problem reduces to an effective $1D$ problem whose Dirac equation
can be written as
\begin{eqnarray} \label{co}
\left(\begin{array}{cc} {V(x)-\varepsilon} & {\frac{d}{dx}+k_y} \\
{-\frac{d}{dx}+k_y} & {V(x)-\varepsilon }
\end{array}\right)\left(\begin{array}{c}
{\phi_{j_{+}}(x)} 
\\ {-i\phi_{j_{-}}(x)}
\end{array}\right)=0.
\end{eqnarray}
An electron which impinges from $x=-\infty$ on the quantum barrier
is partially reflected, partially transmitted at the interface
$x=a$. Inside the barrier regions $[a,c]$ and $[d,f]$, the
eigenstates is a combination of the parabolic cylinder functions
waves. For $x>f$, the carrier is also partly transmitted and
escapes towards $x=\infty$ with a wavevector $+k_{x}$. The
electric potentials $F_{1}$, $F_2$, $F_3$ and $F_4$, being uniform
along the $y$-direction, the $y$-component of momentum is
conserved throughout the {regions}. Due to the space dependence of
the potential $V(x)$ we make the following transformation on our
spinor components to enable us to obtain Schrodinger like
equations for each component,
$\chi_{j_{+}}=\frac{1}{2}\left(\phi_{j_{+}}+\phi_{j_{-}}\right)$
and
$\chi_{j_{-}}=\frac{1}{2i}\left(\phi_{j_{+}}-\phi_{j_{-}}\right)$,
which obey the coupled stationary equations. These are
\begin{eqnarray}\label{eq5}
\frac {d}{dx} \chi_{j_{\pm}}(x)\pm i \left( V(x)-\epsilon
\right)\chi_{j_{\pm}}(x)\mp ik_{{y}}\chi_{{j_{\mp}}}(x) =0.
 \end{eqnarray}
 Each spinor component $\chi_{j_{\pm}}$ can be shown to satisfy
 the following uncoupled second order differential
equation
\begin{equation}\label{df}
\frac {d^{2}}{d{x}^{2}}\chi_{{j_{\pm}}}(x) + \left( \pm i{\frac
{d}{dx}}V(x) + \left[V(x) -\varepsilon \right] ^{2}
-{k_y}^{2}\right) \chi_{{j_{\pm}}}(x)=0.
\end{equation}
At this stage, we point out that our effective $2D$ massless Dirac equation
(\ref{co}) is equivalent to a massive one with an effective mass
$m$ equal to the transverse quantized wave vector $k_y$, i.e. $m=k_y=k_n$. For this
purpose, we consider a unitary transformation, which
enable us to map the effective $2D$ massless equation into a $1D$
massive Dirac equation. Such a unitary transformation does not
affect the energy spectrum or the physics of the problem. We
choose a rotation by $\pi/4$ about the $y$-axis,
$U=e^{i\frac{\pi}{4}\sigma_{y}}$ and thus the transformed
Hamiltonian and wavefunction read
\begin{eqnarray} \label{}
\left(\begin{array}{cc} {V(x)-\varepsilon+k_y} & {\frac{d}{dx}} \\
{-\frac{d}{dx}} & {V(x)-\varepsilon -k_y}
\end{array}\right)\left(\begin{array}{c}
{\tilde{\psi}^{j}_{+}(x)} \\ {} \\ {\tilde{\psi}^{j}_{-}(x)}
\end{array}\right)=0, & \qquad
\tilde{\psi}_{j_{+,-}}(x)=U\psi_{j_{+,-}}(x).
\end{eqnarray}
which is identical to a $1D$ massive Dirac equation with an
effective mass $m^{\ast}=k_y$. This shows clearly how to derive
the dynamical mass generation via space compactification~\cite{Alhaidari2} from our model

\section{Exact solution}

After solving the differential equation (\ref{df}), It turns out
its solution  in regions $x<a$, $[c,d]$ and $x>f$  are given by
\begin{eqnarray}
\phi_{1}(x)&=& \left(%
\begin{array}{c}
  1 \\
  z_{n,k_{x}}\\
\end{array}%
\right)e^{ik_{x}x}+r_{n}\left(%
\begin{array}{c}
  1 \\
  -z^{\ast}_{n,k_{x}} \\
\end{array}%
\right) e^{-ik_{x}x} \\
\phi_{4}(x) &=& \alpha_{n4}\left(%
\begin{array}{c}
  u^{+}_{4}(x) \\
  u^{-}_{4}(x) \\
\end{array}%
\right)+\beta_{n4}\left(%
\begin{array}{c}
v^{+}_{4}(x \\
 v^{-}_{4}(x \\
\end{array}%
\right), \nonumber\\
\phi_{7}(x)&=&t_{n}\left(%
\begin{array}{c}
  1 \\
  z_{n,k_x} \\
\end{array}%
\right)e^{ik_{x}x},
 \end{eqnarray}
where $r_{n}$ and $t_{n}$ are the reflection and transmission
amplitudes, respectively, ${n}$ is {labeling the modes},  the
functions $u^{\pm}_{4}(x)$ and $v^{\pm}_{4}(x)$ are
$u^{+}_{4}(x)=v^{+^{\ast}}_{4}(x)=u^{-}_{4}(x)/z=-v^{-^{\ast}}_{4}(x)=e^{ik_{x}x}$.
The wavevector $k_{x}=\sqrt{\varepsilon^{2}-k_y^{2}}$ and the
complex number $z_{n,k_x}$ is defined as
\beq
 z_{n,k_x}=\mbox{sgn}
(\varepsilon) \frac{k_{x}+ik_{n}} {\sqrt{k_{x}^{2}+k_{n}^{2}}}
\eeq
 where
the transversal momenta is quantized as shown in (\ref{1}). Note
that this quantization is the result of the infinite mass boundary
conditions mentioned previously on the wavefunction along the
$y$-direction. In the case of $|k_{y}|>|\varepsilon|$, the waves
are evanescent (bound states) outside and inside the quantum
barrier and thus the imaginary wavevectors associated with the
evanescent waves are given by
$k_{x}=i\sqrt{k_{y}^{2}-\varepsilon^{2}}$. Since we are interested
by the transmission of relativistic particles (continuum
scattering states), thus we disregard the bound states which
correspond to imaginary $k_{x}$.

The solution of (\ref{df}) in the quantum barrier (region $[a,b]$)
can be expressed in terms of the parabolic cylinder function
$D_{\nu}(x)$ as
\begin{equation}\label{aaaa}
\chi_{2_{+}}(x)=\alpha_{n2}
D_{\nu_{n1}-1}\left(q_{1}\right)+\beta_{n2}
D_{-\nu_{n1}}\left(-q_{1}^{\ast}\right)
\end{equation}
where 
$\nu_{n1}=\frac{ik_{n}^{2}(a-b)}{2V_1}$,
$q_{1}=\sqrt{\frac{2}{(a-b)V_1}}e^{i\pi/4}(V_1 x+E_{1})$,
$E_{1}=-aV_1+(a-b)\varepsilon$, the parameters $\alpha_{n2}$ and
$\beta_{n2}$ are constants. Now substituting (\ref{aaaa}) into
(\ref{eq5})
to get the second component of $\chi_{2}(x)$
\begin{eqnarray}\label{cc}
\chi_{2_{-}}(x)&=&\alpha_{n2}
\frac{-1}{(a-b)k_n}\sqrt{2(a-b)V_1}e^{-i\pi/4}
D_{\nu_{n1}}\left(q_{1}\right) \\
&+&\beta_{n2}
\frac{1}{(a-b)k_n}[-\sqrt{2(a-b)V_1}e^{i\pi/4}D_{1-\nu_{n1}}\left(-q_{1}^{\ast}\right)
2(-V_1 x-E_{1})D_{-\nu_{n1}}\left(-q_{1}^{\ast}\right)]\nonumber.
\end{eqnarray}
The components of the spinor solution of the Dirac equation
(\ref{ak}) in the region $[a,b]$ can be obtained from (\ref{aaaa})
and (\ref{cc}) with $\phi_{2_{+}}(x)=\chi_{2_{+}}+i\chi_{2_{-}}$
and $\phi_{2_{-}}(x)=\chi_{2_{+}} -i\chi_{2_{-}}$. These give
\begin{equation}
\phi_{2}(x)= \alpha_{n2}\left(%
\begin{array}{c}
  u_{2}^{+}(x) \\
  u_{2}^{-}(x) \\
\end{array}%
\right)+\beta_{n2}\left(%
\begin{array}{c}
  v_{2}^{+}(x) \\
  v_{2}^{-}(x) \\
\end{array}%
\right)
\end{equation}
where the functions $u_{2}^{\pm}(x)$ and $v_{2}^{\pm}(x)$  read as
\begin{eqnarray}\label{uu}
u_{2}^{\pm}(x)&=&\mp \sqrt{\frac{2V_1}{a-b}}\frac{1}{k_{n}}e^{i\pi/4}D_{\nu_{n1}}\left(q_{1}\right)+D_{\nu_{n1}-1}\left(q_{1}\right)\nonumber\\
v_{2}^{\pm}(x)&=&\frac{1}{(a-b)^{3/2}k_n}[\pm \sqrt{2V_1}(a-b)e^{-i\pi/4}D_{-\nu_{n1}+1}\left(-q_{1}^{\ast}\right)\\
&+&\sqrt{a-b}\left(b(\pm 2i\varepsilon-k_n)-a(\pm
k_n+2i(V_1-\varepsilon))\mp
2iV_1x\right)D_{-\nu_{n1}}\left(-q_{1}^{\ast}\right)]\nonumber
\end{eqnarray}
Similarly, the solution of (\ref{df}) in the region $[b,c]$ takes
the form
\begin{equation}\label{aaaaa}
\chi_{3_{+}}(x)=\alpha_{n3}
D_{\nu_{n2}}\left(q_{2}\right)+\beta_{n3}
D_{-\nu_{n2}-1}\left(-q_{2}^{\ast}\right)
\end{equation}
where 
$\nu_{n2}=\frac{ik_{n}^{2}(b-c)}{2V_1}$,
$q_{2}=\sqrt{\frac{2}{(b-c)V_1}}e^{i\pi/4}(V_1 x+E_{2})$,
$E_{2}=-cV_1+(c-b)\varepsilon$. The other component of
$\chi_{3}(x)$ is given by
\begin{eqnarray}\label{ccc}
\chi_{3_{-}}(x)&=&\alpha_{n2} \frac{1}{(b-c)k_n}[-\sqrt{2 (b-c)V_1}e^{-i\pi/4}D_{\nu_{2}+1}\left(q_{2}\right)+ 2(V_1 x+E_{2})D_{\nu_{n2}}\left(q_{2}\right)]\nonumber\\
&+&\beta_{n2} \frac{-1}{(b-c)k_n}\sqrt{2 (b-c)V_1}e^{i\pi/4}
D_{-\nu_{n2}}\left(-q_{2}^{\ast}\right).
\end{eqnarray}
Combining \eqref{aaaaa} and \eqref{ccc} in similar way to
${\phi_{2}(x)}$, we obtain the eigenspinor solution of the Dirac
equation (\ref{ak}) in the region $[b,c]$
\begin{equation}
\phi_{3}(x)= \alpha_{n3}\left(%
\begin{array}{c}
  u_{3}^{+}(x) \\
  u_{3}^{-}(x) \\
\end{array}%
\right)+\beta_{n3}\left(%
\begin{array}{c}
  v_{3}^{+}(x) \\
  v_{3}^{-}(x) \\
\end{array}%
\right)
\end{equation}
where we have set
\begin{eqnarray}\label{vv}
u_{3}^{\pm}(x)&=&\frac{1}{(b-c)^{3/2}k_n}[\mp \sqrt{2V_1}(b-c)e^{i\pi/4}D_{\nu_{n2}+1}\left(q_{2}\right)\nonumber\\
&+&\sqrt{b-c}\left(b(\mp 2i\varepsilon+k_n)-c(k_n+2i(\pm V_1\mp\varepsilon))\pm 2iV_1x\right)D_{\nu_{n2}}\left(q_{2}\right)]\\
v_{3}^{\pm}(x)&=&\pm
\sqrt{\frac{2V_1}{b-c}}\frac{1}{k_{n}}e^{-i\pi/4}D_{-\nu_{n2}}\left(-q_{2}^{\ast}\right)+D_{-\nu_{n2}-1}\left(-q_{2}^{\ast}\right)\nonumber
\end{eqnarray}
Finally, note that the general solution of equation (\ref{df}) in
regions $[d,e]$ and $[e,f]$ can be obtained by interchanging
$a\rightarrow d$, $b\rightarrow e$, $c\rightarrow f$ and
$V_1\rightarrow V_2$ in the equations (\ref{uu}) and (\ref{vv}).
{The coefficients} $r_n$, $t_n$, $\alpha_{nj}$ and $\beta_{nj}$
($j=2, 3, 5, 6$)
{can be determined 
by matching wavefunction at different interfaces.}


\section{Transport properties}

The transmission coefficient is determined by imposing the
continuity of the wavefunction at the interfaces between regions.
This procedure is most conveniently expressed in the transfer
matrix formalism. Here we directly use this approach and refer the
reader to references~\cite{BH,M} for a detailed discussion. The
transfer matrix defined by
\begin{eqnarray}\label{}
\hat{M}=\left(
\begin{array}{cc}
  M_{11} &  M_{12} \\
  M_{21} &  M_{21}
\end{array}
\right)
\end{eqnarray}
relates the wavefunction on the left side of the barrier structure
$(\phi_{{x<a}})$ to that 
on the right side
$(\phi_{{x>f}})$. We can then construct  $2\times2$ matrices in
each {region}, whose columns are given by the spinor solutions, such as
for regions (1,4,7)
\begin{eqnarray}\label{}
w_1(x) =w_4(x)=w_7(x)= \left(
\begin{array}{cc}
 e^{i k_{x}x}  & e^{-i k_{x}x} \\
 z_{n,k_{x}} e^{i k_{x}x} & -z^{\ast}_{n,k_{x}} e^{-i k_{x}x}
\end{array}
\right)
\end{eqnarray}
and for $j={2}, 3, 5, 6$
\beq
 w_j(x)=\left(
\begin{array}{cc}
  u_{j}^{+}(x) & v_{j}^{+}(x) \\
 u_{j}^{-}(x) & v_{j}^{-}(x)
\end{array}
\right)
\eeq
These matrices play the role of partial
transfer matrices and allow  to express the continuity condition
 of the wavefunction at each {interface}. Note that the
components of matrices $w_5(x)$ and $w_6(x)$ can be obtained by
interchanging $a\rightarrow d$, $b\rightarrow e$ and $c\rightarrow
f$ in matrices $w_2(x)$ and $w_3(x)$, respectively. After
straightforward algebra, we get the transfer matrix as function of
different boundaries
\begin{eqnarray}\label{}
\hat{M}=w_1^{-1}(a)  w_2(a)  w_2^{-1}(b)  w_3(b)  w_3^{-1}(c)
w_1(c) w_1^{-1}(d)  w_5(d)  w_5^{-1}(e)  w_6(e)  w_6^{-1}(f)
w_1(f)
\end{eqnarray}
and the relation which expresses the continuity of the
wavefunction is then given by
\begin{eqnarray}\label{rt}
\left(
  \begin{array}{c}
    1 \\
    r_{n} \\
  \end{array}
\right) =\hat{M} \left(
           \begin{array}{c}
             t_{n} \\
             0 \\
           \end{array}
         \right).
\end{eqnarray}
Solving equation (\ref{rt}) for the transmission amplitude $t_{n}$
of the $n$th wave mode through the barrier, we get the
transmission probability $T_{n}$ as
\begin{eqnarray}\label{}
T_{n}=\left|t_{n}\right|^{2}=\frac{1}{\left|M_{11}\right|^{2}}.
\end{eqnarray}

Based on \cite{danneau}, we
give a review about the shot noise.
Indeed, the conductance of a single transmission channel can be written as
\beq G=g\frac{e^{2}}{h}T \eeq
 where $g$ is the degeneracy (spin and
valley) of the system and $T$ the electron transmission
probability. When the system is biased, shot noise appears due to
discreteness of charge~\cite{blanter} and these current
fluctuations for a single channel are given by \beq \langle(\delta
I)^{2}\rangle=2e\langle I\rangle(1-T). \eeq The total noise power
spectrum for a multichannel conductor is then obtained by summing
over all $N$ transmission eigenchannels:
\begin{eqnarray}\label{}
S_I=\frac{2e^{3}|V|}{h}\sum^{N_{\sf{max}}-1}_{n=0}T_{n}(1-T_{n}).
\end{eqnarray}
In the limit of low transparency $T_n\ll 1$,
\begin{eqnarray}\label{}
S_I\cong S_{\sf
Poisson}=\frac{2e^{3}|V|}{h}\sum^{N-1}_{n=0}T_{n}=2e\langle
I\rangle
\end{eqnarray}
defining a Poissonian noise induced by independent and random
electrons like in tunnel junctions~\cite{blanter}. The regular way
to quantify shot noise is to use the Fano factor $F$ which is the
ratio between the measured shot noise and the Poissonian noise:
\begin{eqnarray}\label{}
F=\frac{S_I}{S_{\sf Poisson}}=\frac{S_I}{2e\langle
I\rangle}=\frac{\sum^{N-1}_{n=0}T_{n}(1-T_{n})}{\sum^{N-1}_{n=0}T_{n}}.
\end{eqnarray}
Then, for a Poissonian process $F=1$ at small transparency
$(T_n\rightarrow0)$, while $F=0$ in the ballistic regime (i.e.
when $T_n\rightarrow1$) and $F=1/3$ in the case of a diffusive
system~\cite{Tworzydle, Snyman, Peres, DiCarlo}.

In graphene, it has been theoretically concluded that transport at
the Dirac point occurs via electronic evanescent
waves~\cite{Katsnelson,Tworzydle}. Tworzydlo {\it et
al.}~\cite{Tworzydle} used heavily-doped graphene leads and the
wavefunction matching method to directly solve the Dirac equation
in perfect graphene with
length $L_B$ and width $w$. 
They found that for armchair edges, the quantization condition of
the transverse wave vector is defined by \beq
k_{y,n}=\frac{(n+\alpha)}{w}\pi \eeq where $\alpha=0$ or $1/3$ for
metallic and semiconducting armchair edges, respectively. At the
Dirac point, the transmission coefficients are given
by~\cite{Tworzydle}
\begin{eqnarray}\label{}
T_n=\frac{1}{\cosh\left(\pi\left(n+\alpha\right)\frac{L_B}{w}\right)}.
\end{eqnarray}
Consequently, graphene has a similar bimodal distribution of
transmission eigenvalues at the Dirac point as there is in
diffusive systems~\cite{benn, Nagaev}. In the limit of
$w/L_B\rightarrow\infty$, the mode spacing becoming small and one
can replace the sum over the channels by an integral over the
transverse wave vector component $k_y$ to obtain the conductivity
and the Fano factor for a sheet with metallic armchair
edge~\cite{Tworzydle}
\begin{eqnarray}
\sigma_{\sf
Dirac}&=&G\frac{L_B}{w}=\frac{4e^{2}}{h}\frac{L_B}{w}\int^{\infty}_{0}
\frac{dk_y}{\cosh^{2}(k_yL_B)}=\frac{4e^{2}}{\pi h}\label{sssss}\\
F_{\sf
Dirac}&=&\frac{\sum^{N-1}_{n=0}T_{n}(1-T_{n})}{\sum^{N-1}_{n=0}T_{n}}=\frac{1}{3}\label{fffff}.
\end{eqnarray}

In summary, we will study the above quantities for the present system in terms of our findings
and compare with already published works. In fact,
%
 the conductivity and the Fano factor of  Dirac fermions  through  one and
double triangular barriers in graphene will have a variate and
different from with respect to the results presented in
\eqref{sssss} and \eqref{fffff}.

\section{Results and discussions}

For a better understanding of the obtained results so far,
we numerically analysis different physical quantities in terms of the
system parameters. To underline their behaviors,
we trait
single and double triangular barriers, separately.

\subsection{Single barrier}
We start our discussion by studying the transmission probability
and  shot noise for the Dirac fermions scattered by a single
triangular barrier potential. We implement our previous analytical
approaches to a graphene system subject to a single triangular
barrier potential of strength $V_1$ and $V_2=0$. We will see that
the transmission coefficient has a rich information about the
electronic transport properties of the Dirac fermions through a
triangular barrier structure.

\begin{figure}[H]
  \begin{center}
  \includegraphics[width=2.8in]{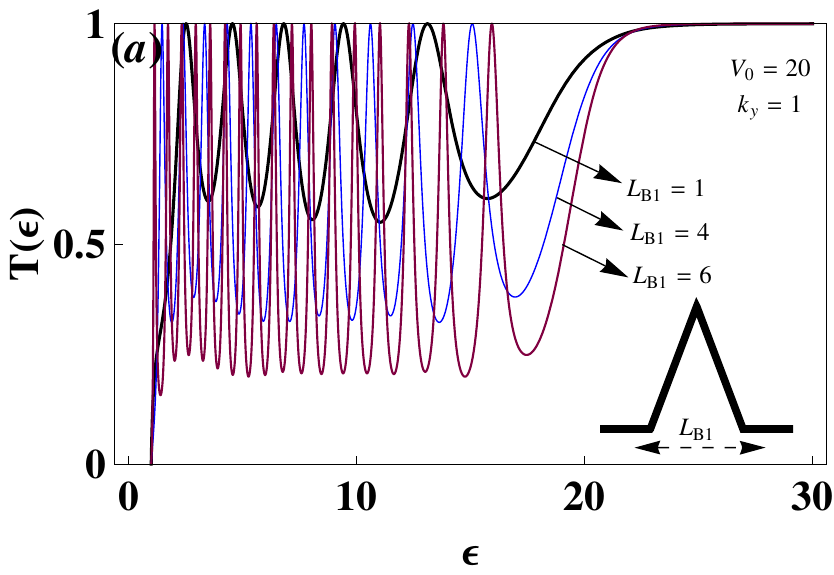}\ \ \ \
    \includegraphics[width=2.8in]{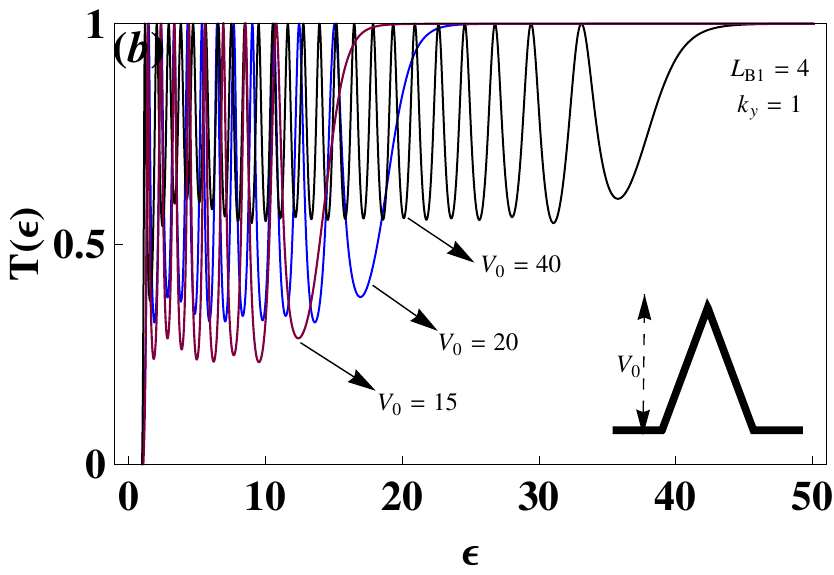}\\
  \includegraphics[width=2.8in]{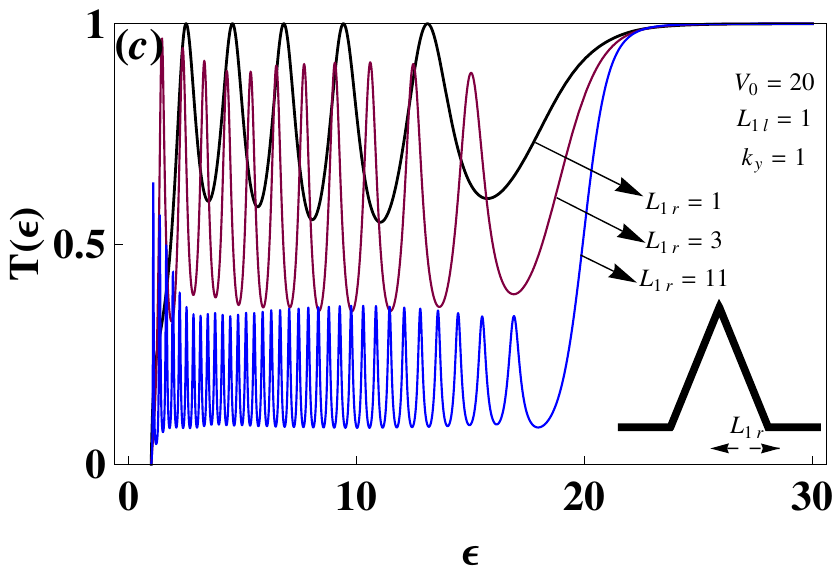}\  \ \ \
  \includegraphics[width=2.8in]{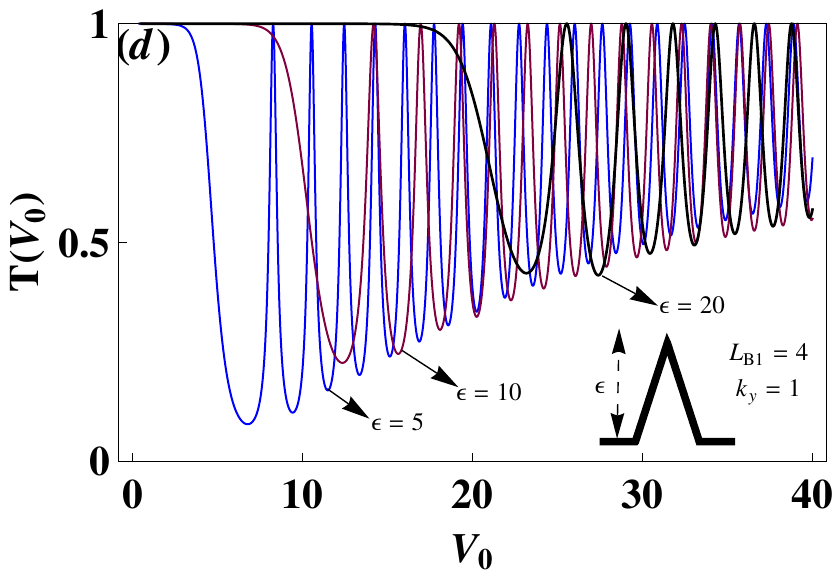}
  \end{center}
   \caption{\sf{Transmission coefficient $T$ for the Dirac fermion
  scattered by a single triangular barrier potential with
 $V_1=V_0$ and $V_2=0$. (a), (b) and (c) as a function of incident
 electron energy, (d) as a function of
applied voltage.  {"Color figure online"}}}\label{fig2}
\end{figure}

The variations of the calculated transmission coefficient $T$ in
terms of the incident electron energy $\varepsilon$ and applied
voltage
$V_0$ is displayed in Figure \ref{fig2} 
for  different values of 
the barrier widths $L_{B1}$, 
barrier widths
right side $L_{1r}$, 
barrier heights $V_0$ 
and
incident electron energy $\varepsilon$. 
From Figure \ref{fig2}, one can see that for certain values of
$L_{B1}$ (Figure \ref{fig2}a), $V_0$ (Figure \ref{fig2}b) and
$L_{1r}$ (Figure \ref{fig2}c), the transmission resonances appear
in the triangular potential for
$m^{\ast}<\varepsilon<V_0+2m^{\ast}$ and vanishes for both
conditions ($\varepsilon>V_0$, $L_{1r}\neq
L_{1l}$). We note that 
the intensity of resonances increase as long as 
$V_0$ and 
$L_{B1}$ increase, which allow for emergence of
peaks in the $T$ shape. 
One can {see} that $T$ 
always starts from the energy  corresponding to $k_y=m^{\ast}$,
with $m^{\ast}$ is the effective mass of the $1D$ Dirac fermion.
The zone when we have the energy such as $\varepsilon<m^{\ast}$
corresponds to the forbidden zone. It is important to note that
the resonant energy  depends strongly on the barrier height and
width.
For $L_{B1}=4$ we have the transmission resonances independtly of
the value taken by the applied potentail $V_0$ as long as
$V_0<\varepsilon$.
While for $V_0>\varepsilon-2m^{\ast}$, the resonances decrease
sharply  until reach a relative minimum and then begin to increase
in an oscillatory manner.

\begin{figure}[h!]
\centering{
   \includegraphics[width=2.8in]{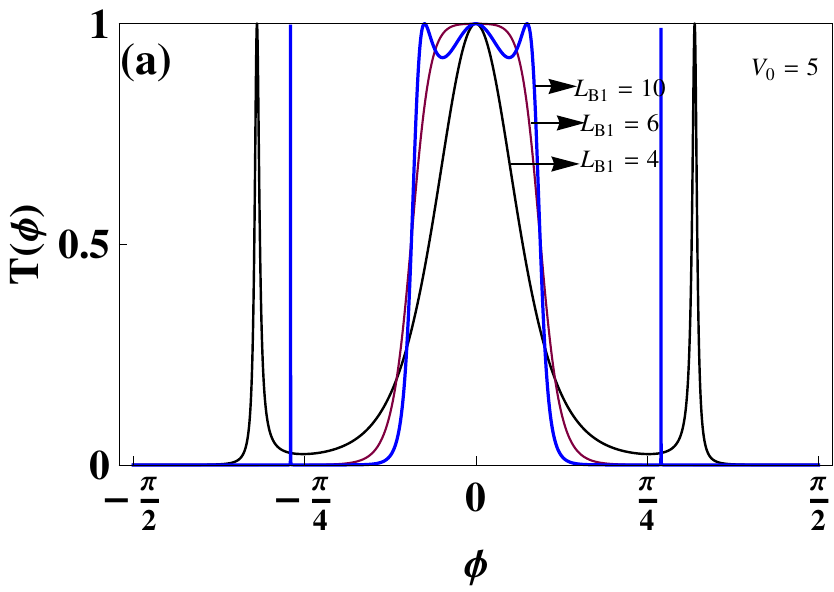}\\
   \includegraphics[width=2.8in]{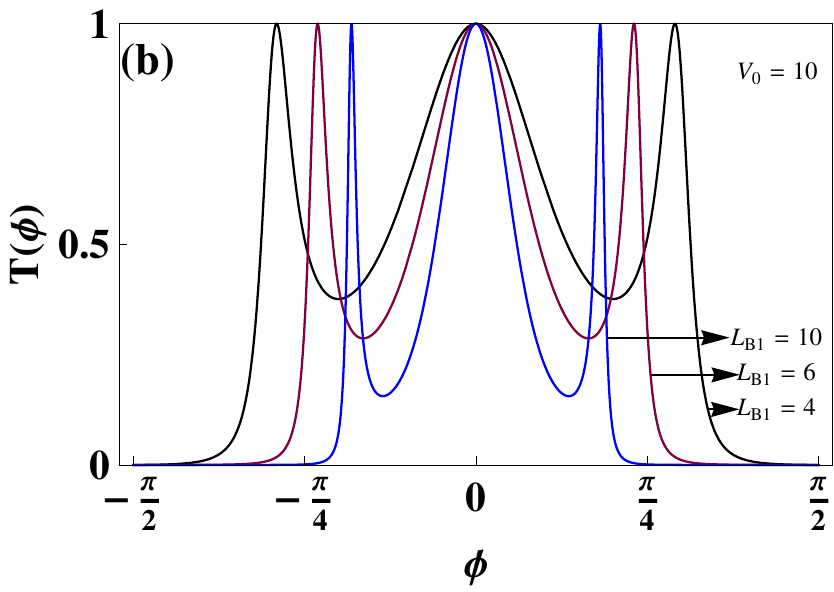}\ \ \ \ \ \ \ \ \ \ \ \
  \includegraphics[width=2.8in]{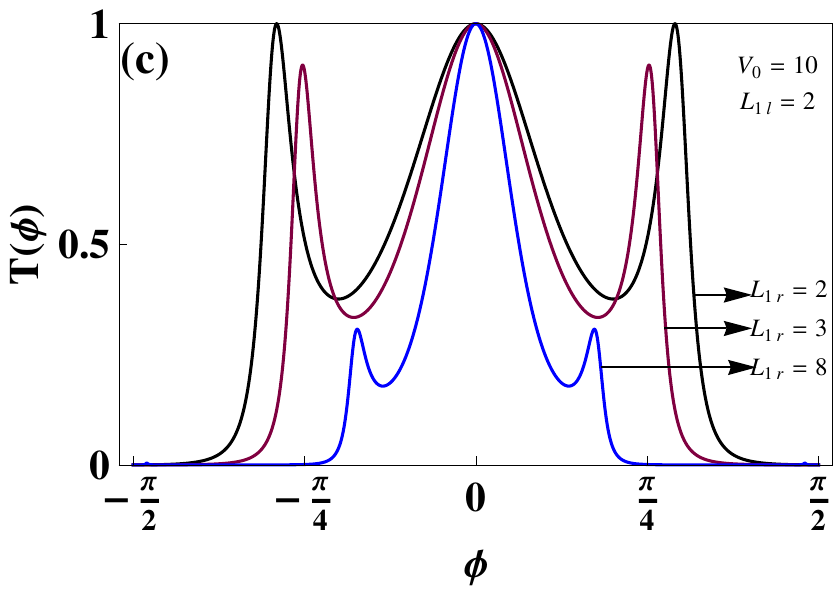}}\\
   \caption{\sf{Transmission coefficient $T$ as a function of electron incident angle $\phi$
    for the Dirac fermion scattered by a single triangular barrier
  potential with $V_1=V_0$, $V_2=0$
  and $\varepsilon=2$. "Color figure online"}}\label{fig3}
\end{figure}

Figure \ref{fig3} is showing the transmission coefficient $T$ as
function of the electron incident angle $\phi$
for $\varepsilon=2$ and different values of
($V_0$, 
$L_{B1}$,
$L_{1r}$). We see that 
the perfect transmission occurs at different angles and vice
versa. It is observed that, the transmission is always total for a
normal
incidence angle. 
For $V_0=5$ one can observe that $T$  is not zero
for some values of the barrier width. 
In particular it shows up two peaks at incident angles
$\phi=\pm\frac{\pi}{3.6877}$ and $\phi=\pm\frac{\pi}{3.1469}$ for
each value of the barrier widths $L_{B1}=10$ and $L_{B1}=4$,
respectively. The transmission resonances  always appear for the
case of the barrier width only of the right side $L_{1r}$ is equal
to the barrier width only of the left side $L_{1l}$, i.e.
$L_{1r}=L_{1l}$,
while 
disappear otherwise $(L_{1r}\neq L_{1l})$.\\

\begin{figure}[h!]
\centering
  \includegraphics[width=3.2in]{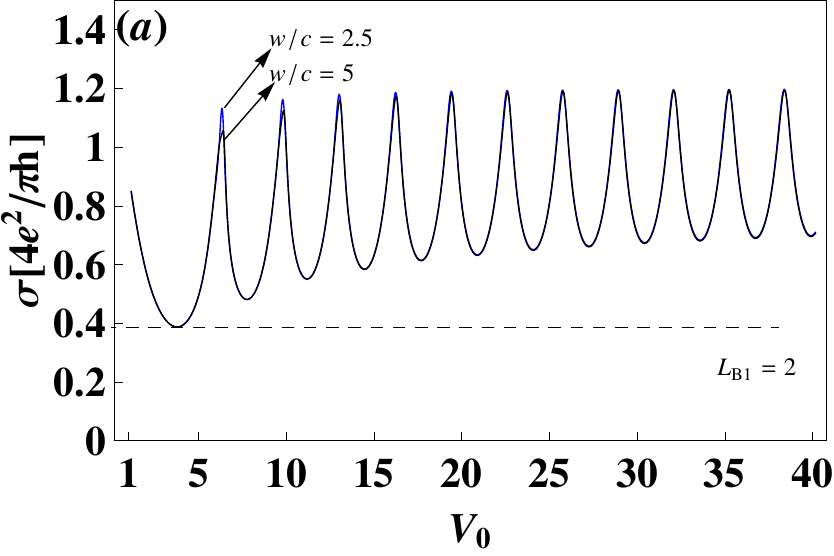}\\
  \includegraphics[width=3.2in]{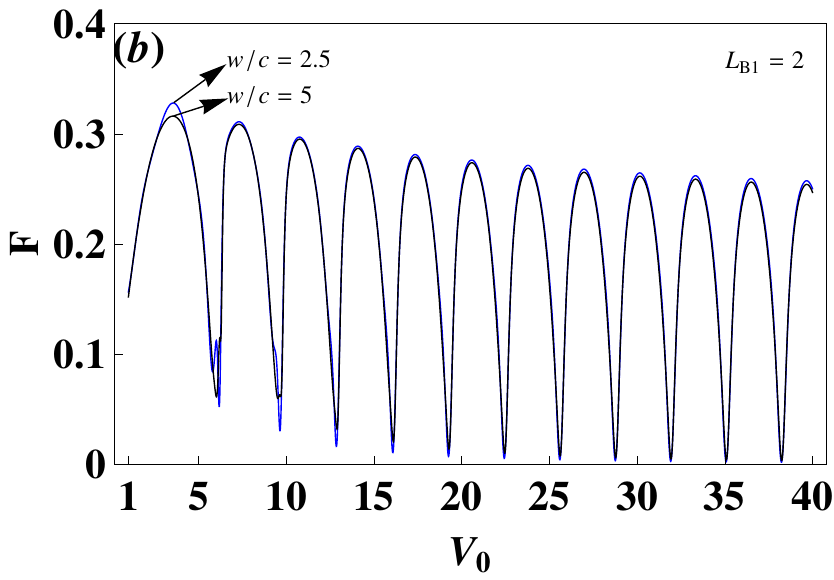}\\
   \caption{\sf{The electrostatic potential dependence of
  the Fano factor and the conductivity for the Dirac fermion scattered by a
  single triangular barrier potential  with $V_1=V_0$, $V_2=0$,
  and $\varepsilon=2$. "Color figure online"}}\label{fig4}
\end{figure}

In what follows we discuss 
 the conductivity $\si$ and Fano factor $F$ behaviors to underline what makes difference
 with \cite{Gusynin,Tworzydle,Gusyninn}. Indeed in
Figure \ref{fig4},
\noindent we plot 
$\sigma$ (in units of $4e^{2}/\pi h$) and
$F$ in terms of the
electrostatic potential $V_0$ for 
$L_{B1}=2$ and  
$\varepsilon=2$. It is interesting to note that $\si$
corresponding to our system
is showing some differences with respect to that for an ideal
strip of graphene \cite{Tworzydle}, which supports perfect
transmission regardless of the barrier height (Klein
tunneling~\cite{Katsnelsonn}).
It is obvious to observe that the
conductivity and Fano factor change their behavior in an
oscillatory way (irregularly periodical tunneling peaks) when we
augment the potential of applied voltage. One can see that as long
as $V_0$ increases,
the number of minimum of $\si$
increases as well but the associated
number
of maximum of $F$
decreases. This effect is  different from that obtained
in~\cite{Tworzydle} where there is only one minimum conductance
$4e^{2}/(\pi h)$ at the Dirac point and for a geometric factor
$w/L_B=5$, which corresponds to one maximum Fano factor $1/3$. In
contrast, in our case for the same factor ($w/L_B=5$) the minimum
conductivity  0.387 appears around $V_0=3.661$ where the
associated maximum Fano factor is
0.315.
More importantly, for two values of $V_0$ like $3.661$ and $7.616$
one can see that the minimum conductivity increases from  0.387 to
0.479.
{Consequently, both the potential barrier height and width for the particles emission can be reduced
 and then they can easily tunnel through the full barrier width, causing a larger field emission current}.
Therefore, we conclude that it is relatively more easily for the
Dirac fermions to tunnel through a triangular barrier in a
graphene sheet
rather than rectangular one. It should be pointed out that the
nonzero minimum conductance, as shown in Figure \ref{fig4}, may
due to the conservation of pseudospin and the chiral nature of the
relativistic particles in the graphene nanoribbon.

\subsection{Double barriers}

\begin{figure}[h!]
\centering
  \includegraphics[width=2.8in]{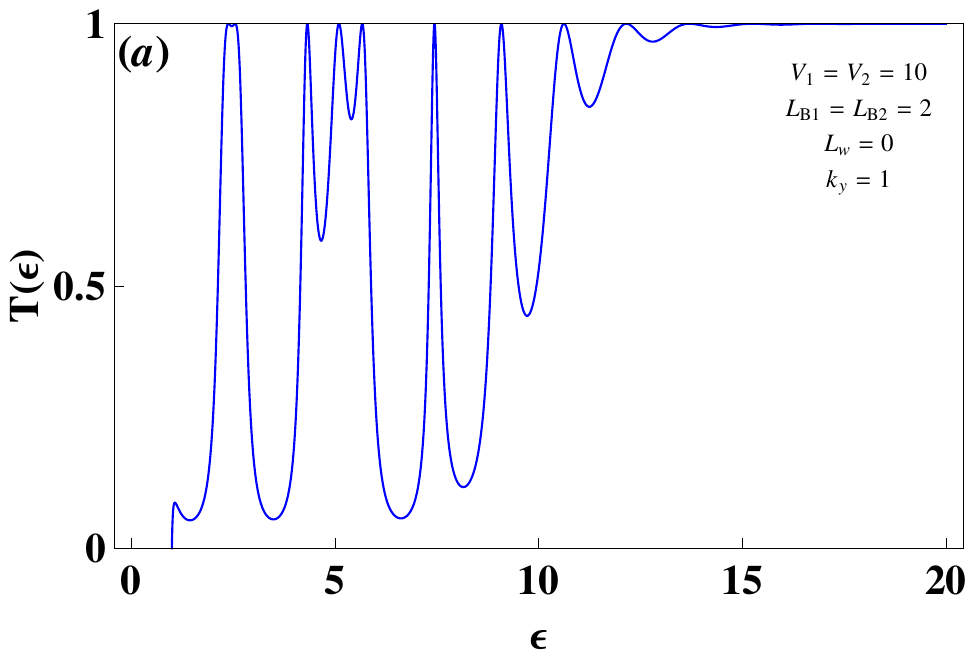}\ \ \ \ \ \ \ \ \ \ \ \
  \includegraphics[width=2.8in]{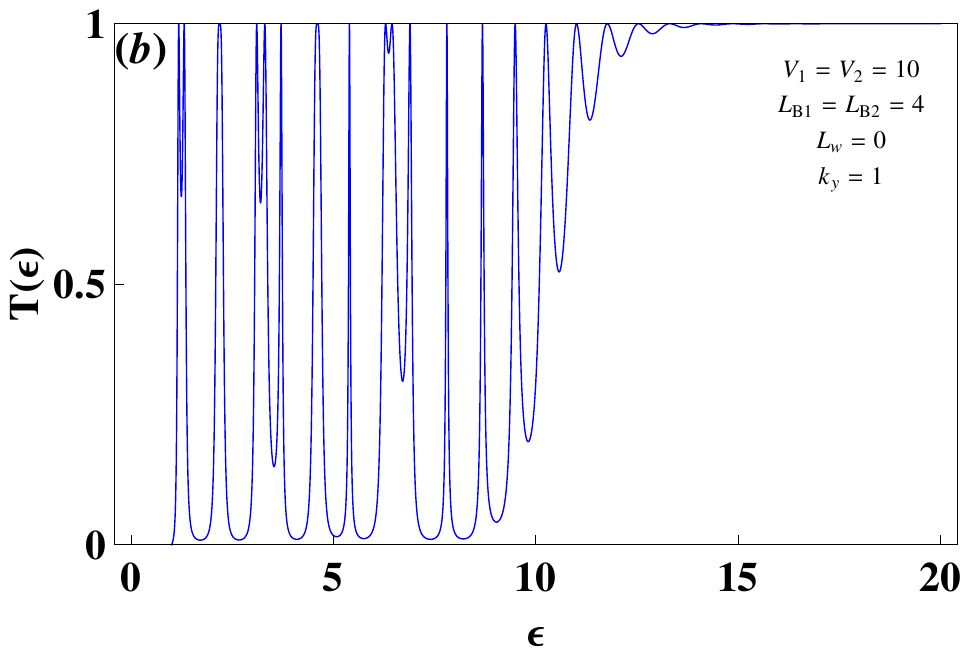}\\
  \includegraphics[width=2.8in]{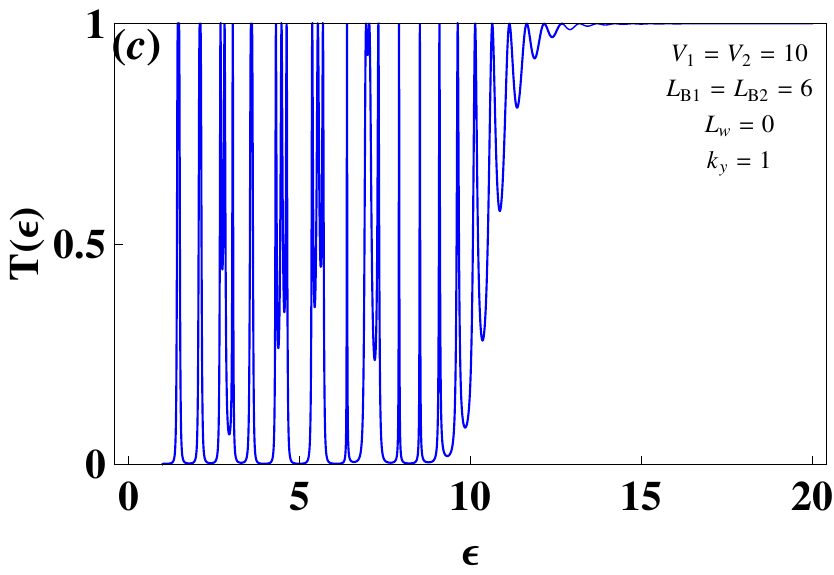}\ \ \ \ \ \ \ \ \ \ \ \
  \includegraphics[width=2.8in]{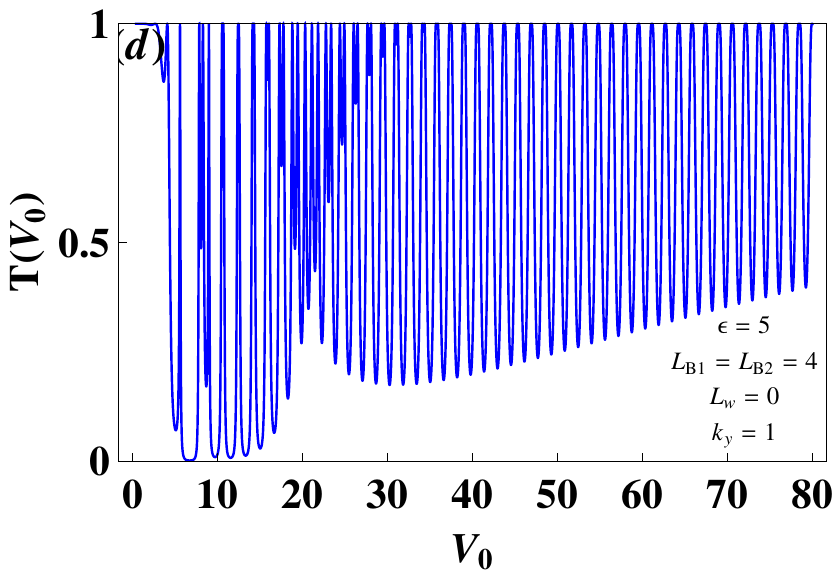}\\
  \includegraphics[width=2.8in]{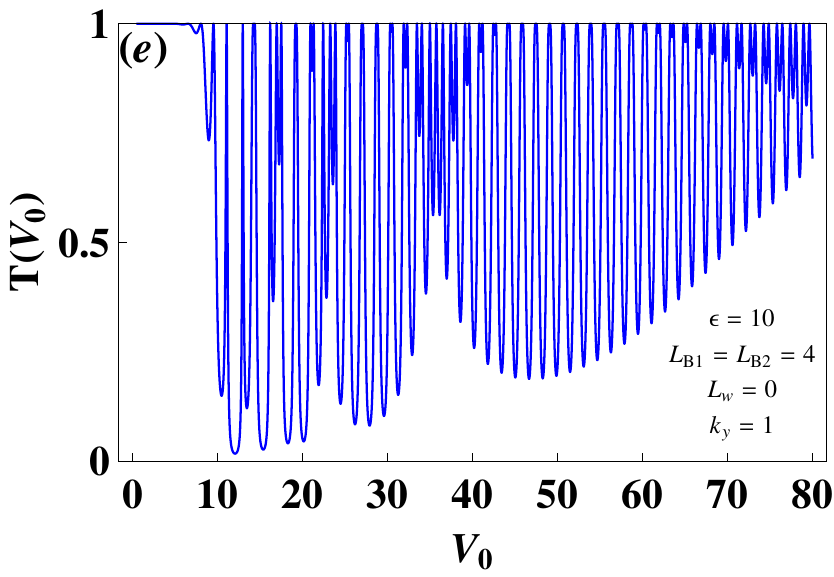}\ \ \ \ \ \ \ \ \ \ \ \
  \includegraphics[width=2.8in]{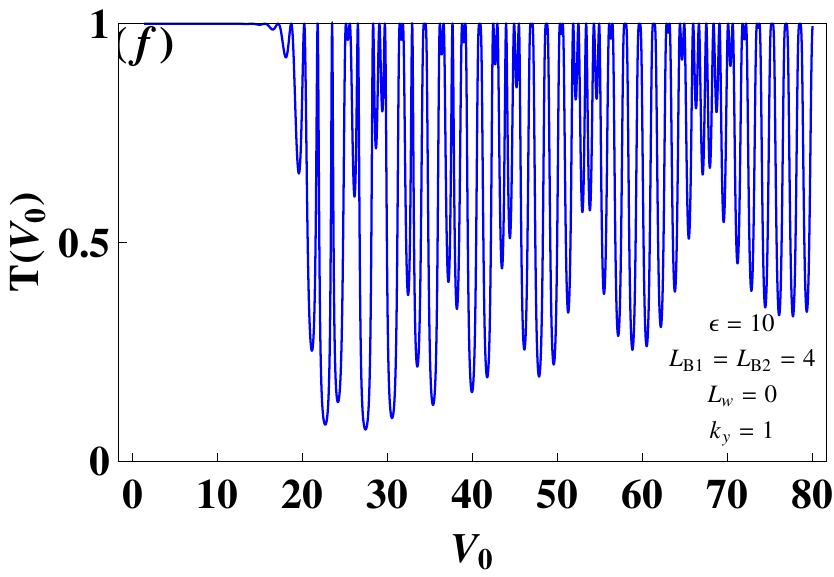}\\
   \caption{\sf{Transmission coefficient $T$ for the Dirac fermion scattered by a double triangular barriers potential
with $V_1=V_2=V_0$, $L_w=0$ and $L_{B1}=L_{B2}$. (a), (b) and (c)
as a function of incident electron energy, (d), (e) and (f) as a
function of applied voltage. "Color figure online"}}\label{fig5}
\end{figure}

\begin{figure}[H]
\centering
  \includegraphics[width=3.2in]{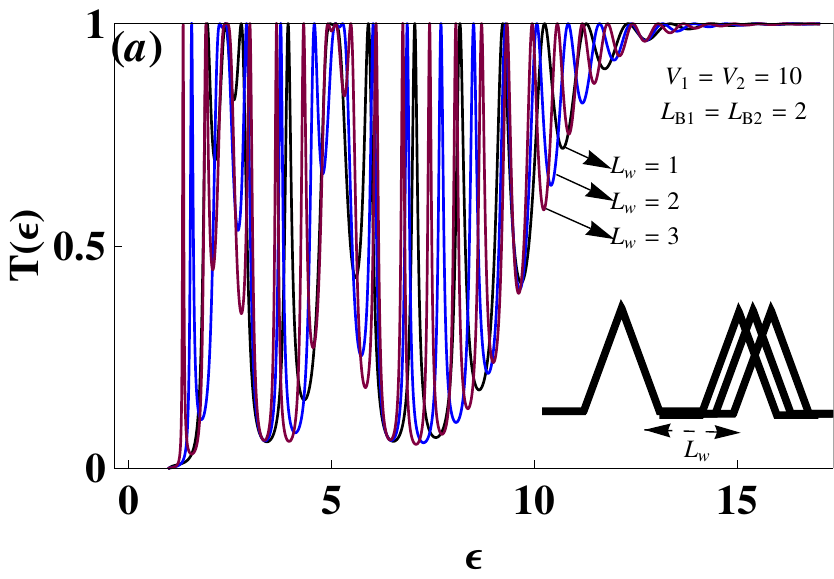}\ \ \ \
  \includegraphics[width=3.2in]{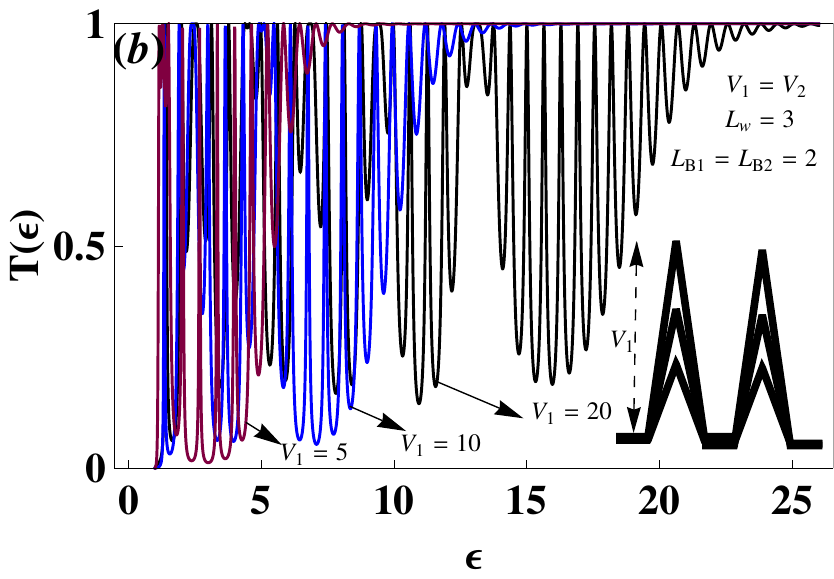}\\
  \includegraphics[width=3.2in]{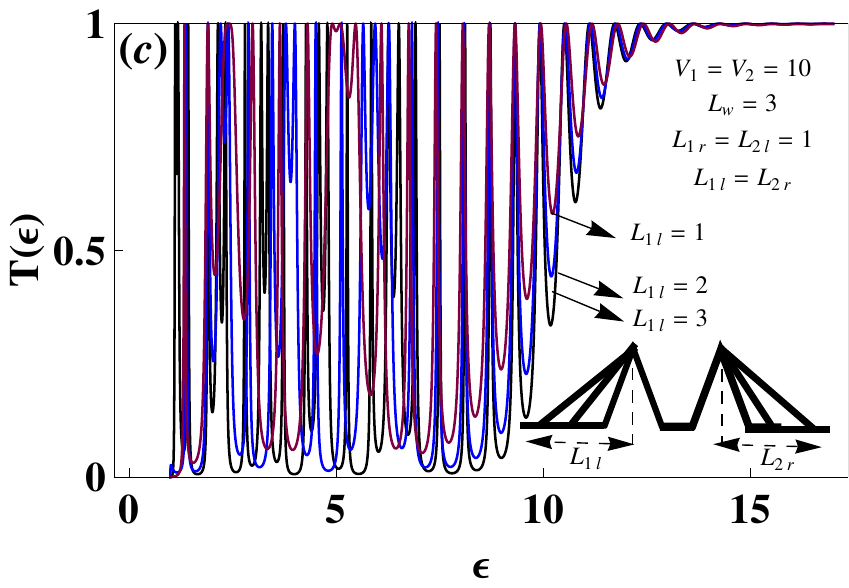}\ \ \ \
  \includegraphics[width=3.2in]{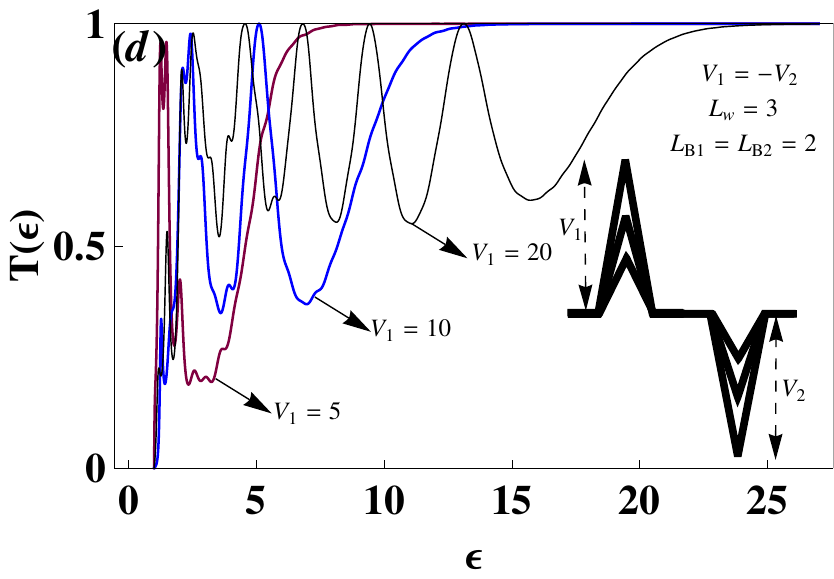}\\
   \caption{\sf{Transmission coefficient $T$ as a function of incident electron energy for the Dirac fermion
scattered by a double triangular barriers potential with
$V_1=|V_2|$ and $k_y=1$. (a) the width of the well was varied, (b)
the barrier height is varied, (c) the barrier width only of the
outer sides is varied, (d) the height of the barrier and the well
is varied. {"Color figure online"}}}\label{fig8}
\end{figure}

In this section we implement our previous analytical approach to a
graphene system subject to a double triangular barrier potentials
 $V_1$ and $V_2$ so that the resulting static electric
field strengths are
{\beq
F_1=\frac{V_1}{l_{1l}}, \qquad F_2=-\frac{V_1}{l_{1r}}, \qquad
F_3=\frac{V_2}{l_{2l}}, \qquad F_4=-\frac{V_2}{l_{2r}}.
\eeq}
Note that there are various parameters involved 
such as well width, barrier height and barrier width, these will
offer different discussions about transport properties in the
present configuration of potential.
In particular, Figures \ref{fig5} and \ref{fig8} show the
transmission coefficient
in terms of
 the incident electron  energy $\varepsilon$ and applied
voltage  $V_1=V_2=V_0$ 
for both cases $L_w=0$ and $L_w\neq0$, with $L_w$ is
the interbarrier separation (well width). 
In Figure \ref{fig5}, one can see that contrary to single barrier
(e.g. Figure 2c) the transmission resonances always appear for
 double triangular barrier case.  Clearly, the
intensity and width of resonances as well as the condition for the
existence of resonances depend on the static electric field
strengths ($F_1$, $F_2$, $F_3$, $F_4$) and
barrier widths ($L_{B1}$, $L_{B2}$). The intensity of resonances
increases and decreases as long as
the strengths ($\left|F_1\right|$, $\left|F_2\right|$,
$\left|F_3\right|$, $\left|F_4\right|$) decrease and  widths
($L_{B1}$, $L_{B2}$) increase, respectively.\\

\noindent In Figure \ref{fig8} we consider the same behavior of transmission
as before but with  $L_{w}\neq0$.
Compared to Figure
\ref{fig5} 
for  $V_1=V_2=10$,
one can conclude that 
the intensity of
resonances depends strongly on 
$L_{w}$.
 In addition for
$V_1=V_2$ there are several peaks showing transmission resonances
those correspond to the bound states and no resonances exist for
$V_1=-V_2$. Figure \ref{fig8}a presents three potential profiles
and
 their transmission coefficients in terms of $\varepsilon$ 
for   $L_w=1,2,3$, $L_{B1}=L_{B2}=2$ and $V_1=V_2=10$. The results
show
that as long as the well width $L_w$ increases 
the resonance peak
structures become sharpened.
We consider in Figure \ref{fig8}b the case of three values of
barrier height 
($V_1=V_2=5, 10, 20$) 
for $L_{B1}=L_{B2}=2$ and $L_w=3$. As
the barrier height is enhanced, the transmission resonance shifts 
and the width of the resonances increases. The case of the double
barriers is more important when the outer slopes of the barriers
are varied with a fixed barrier height as shown in Figure
\ref{fig8}c. We notice that the well region and the inner slope
region are not changed when the parameters ($L_{1l}= L_{2r}$)
vary.
As long as the barrier becomes thick and high the resonance shifts
toward
the higher and the peak is more sharpened. Similarly to
one triangular barrier, no resonance exists for the incident
electron energy higher than
$V_1+2m^{\ast}$ and the zone 
$\varepsilon<m^{\ast}$ is a forbidden zone. Except that
when  $V_1=-V_2$ no resonances exist for the following cases 
($1<\varepsilon<9, V_1=-V_2=5$), ($1<\varepsilon<5, V_1=-V_2=10$)
and ($1<\varepsilon<4.5, V_1=-V_2=20$), which are clearly
 shown in Figure \ref{fig8}d.

\begin{figure}[H]
\centering
   \includegraphics[width=2.2in]{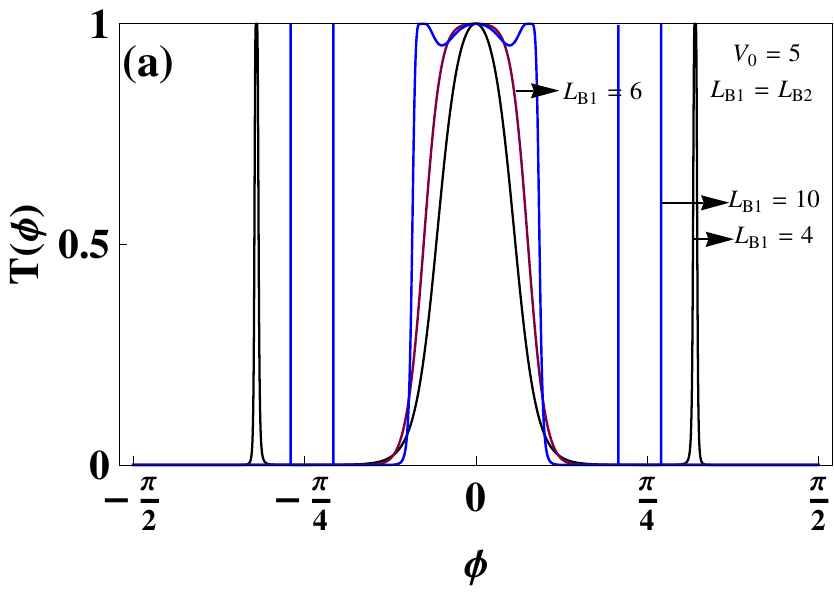}
   \includegraphics[width=2.2in]{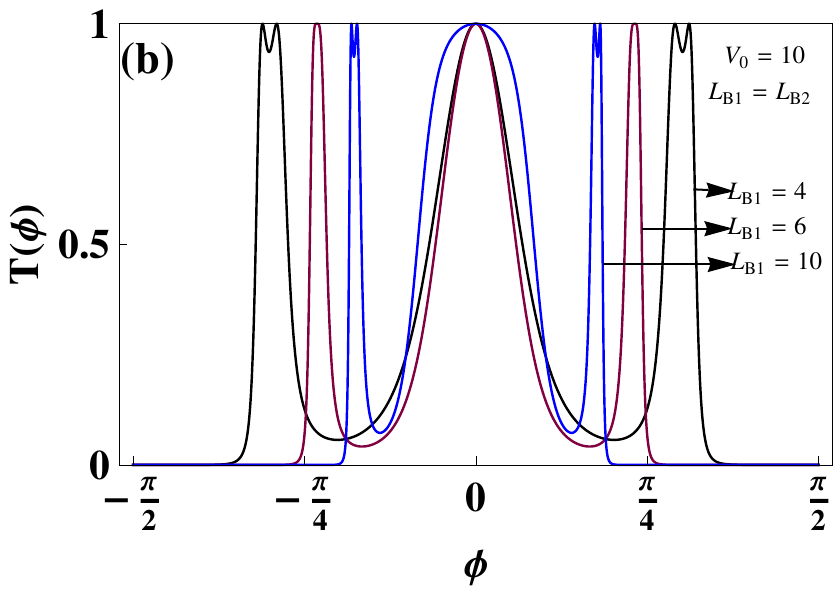}
    \includegraphics[width=2.2in]{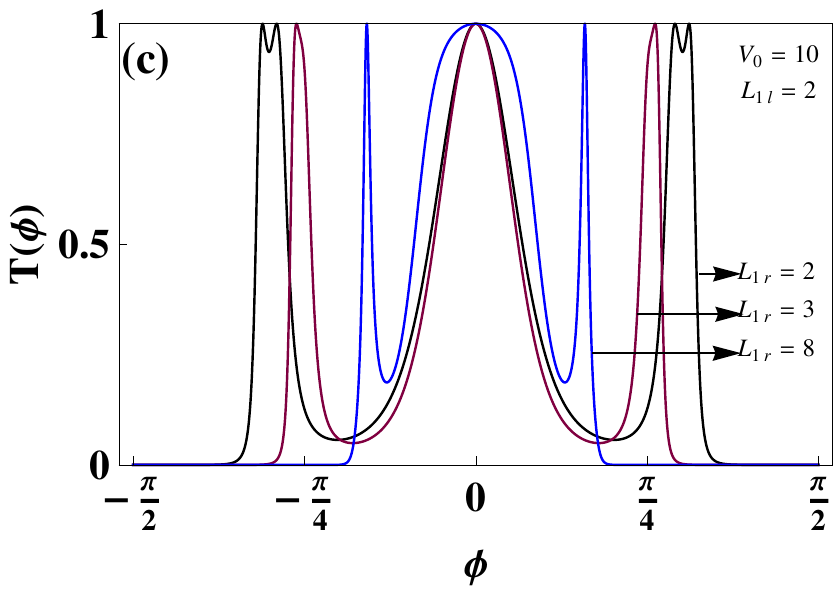}\\
   \caption{\sf{Transmission coefficient $T(\phi)$ for the Dirac fermion
   scattered by a double triangular barriers
potential with $V_1=V_2=V_0$, $L_w=0$ and
$\varepsilon=2$. {"Color figure online"}}}\label{fig6}
\end{figure}

We represent in Figure \ref{fig6} the transmission coefficient
versus the incident angle with the same parameters as in Figure
\ref{fig3} for the Dirac fermion scattered by  double triangular
barriers potential with the interbarrier separation $L_w=0$. By
contrast with the case for the Dirac fermion scattered by a
single triangular barrier potential we conclude that 
the transmission resonances still always exist. The comparison
between these two types of potential shows that for double
barriers of strength $V_1=V_2=5$ we have three peaks with two
peaks at incident angles
\beq
\phi=\pm\frac{\pi}{4.8021}, \qquad
\phi=\pm\frac{\pi}{3.6877}
\eeq
 and one peak at
 \beq
 \phi=\pm\pi
 \eeq
 for
each value of the barrier widths $L_{B1}=10$ and $L_{B1}=4$,
respectively. Even though the barrier width only of the right side
$L_{1r}$ is different to the barrier width only of the left side
$L_{1l}$ for double  barriers, the transmission resonances always
appear contrary to the Dirac fermion scattered by a single
barrier. We observe that decreasing the barrier width only of the
right side the transmission coefficient takes relevant values for
a wider set of incident angles.\\

{Finally, we close our discussion about transmission resonances by making comparison with the results reported
for double barrier in graphene subjected to an external magnetic field in \cite{Massir}. In fact, it was shown that
increasing the magnetic field leads to a shift of transmission cone (a reduction
of the number of resonances) and a shrinking of the perfect transmission region.
However in our present study as we noticed before, the intensity of resonances increases and decreases as long as
 the static electric field strengths
$(|F_1|, |F_2|, |F_3|, |F_4|)$ decrease and barrier widths $(L_{B_1}, L_{B_2})$ increase, respectively.}

\begin{figure} [H]
\centering
  \includegraphics[width=3.2in]{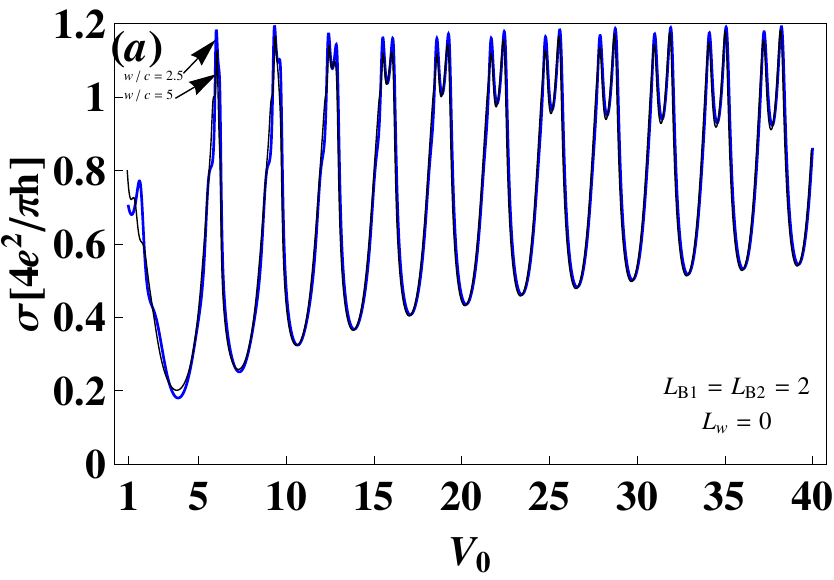} \ \ \ \
  \includegraphics[width=3.2in]{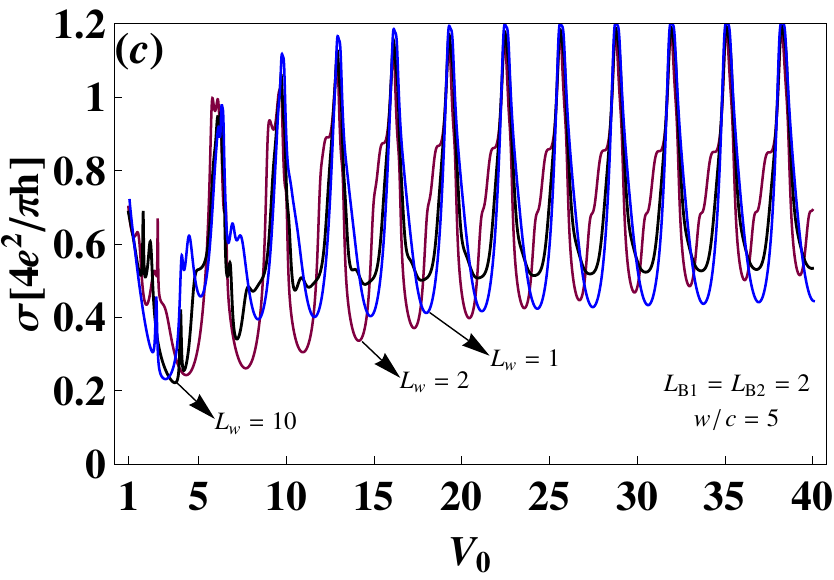}
  \includegraphics[width=3.2in]{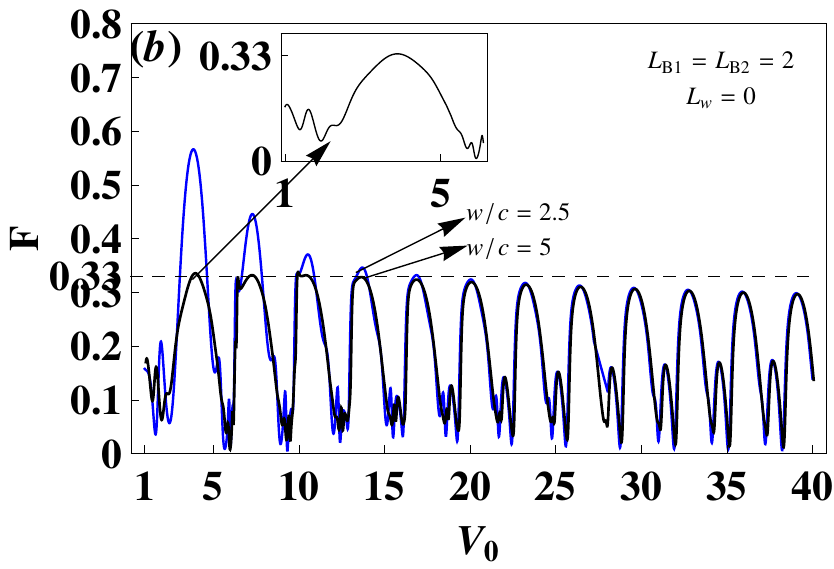}\ \  \ \
  \includegraphics[width=3.2in]{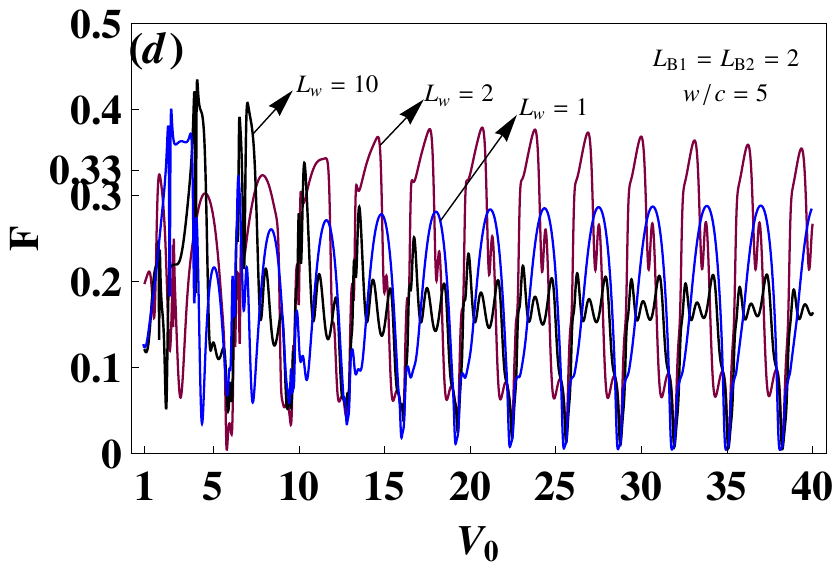}\\
   \caption{\sf{The electrostatic potential dependence of the Fano factor and the conductivity for the Dirac
fermion scattered by a double triangular barriers potential with
$V_1=V_2=V_0$, and $\varepsilon=2$. {"Color figure online"}}}\label{fig7}
\end{figure}

As far as the conductivity and Fano factor
behaviors  for 
double barriers are concerned, we notice that the shot noise is
characterized by the maximum of peaks at the minimums of conductivity 
and  minimum of valleys at the maximums of conductivity. 
The role of the interbarrier separation $L_w$ resulted in
increasing peaks of  shot noise  and lowering  the current of
valleys as shown in Figure \ref{fig7}. 
One can see that the value $F=1/3$ for the Dirac fermion scattered
by  double barriers  is reproduced in the case where the barrier
widths $L_{B1}=L_{B2}=2$, the interbarrier separation $L_w=0$ and
the applied voltage $V_0$ near the $2\varepsilon$.

\section{Conclusion}

We have analyzed the behavior of Dirac fermions in graphene
submitted to electrostatic potential of triangular type. By
solving the eigenvalue equation we have obtained the solutions of
the energy spectrum in terms of different physical parameters
involved in the Hamiltonian system.
Using 
the continuity of the wavefunctions at the interfaces between
regions inside and outside the barriers, we have studied the
transport properties of the present system. More precisely, using
the transfer matrix method, we have 
analyzed the corresponding
transmission coefficient, conductivity and Fano factor 
for single and double triangular barriers.

It has been shown that the Dirac fermions scattered by single
triangular and double triangular barriers own a minimum
conductivity associated with a maximum Fano factor. We have
noticed that the Dirac fermions can tunnel more easily through a
barrier in the triangular forms rather than in the rectangular
one. On the other hand, the
behavior of the conductivity and the Fano factor in terms of 
the applied voltage showed irregular periodical oscillating for
triangular potential.

We have noticed that the resonant energy is influenced  by the barrier width. 
Indeed, when the barrier becomes thick and/or high, the resonant
peak becomes sharpened and shifted to the higher energy. Even if
the thickness and the height of the barriers are constant, the
form of the triangular barrier affects the resonant energy. On the
contrary, the triangular double barriers structures is less
sensitive to the well width compared with a rectangular double
barrier structure. Therefore, we have concluded that it is
relatively more easily for the Dirac fermions to tunnel through a
triangular barrier in a graphene sheet rather than rectangular
one.
{These results may be helpful to deeply understand the transport
in the nanoribbons and design the graphene-based nanodevices}.

We close by mentioning that the obtained results can be extended
to deal with other issues related to graphene systems. Indeed, one
may think to study the transport properties of Dirac fermions
scattered by periodical potentials and other types. Another
interesting problem, what about a generalization of the obtained
results to bilayer graphene and related matter.

\section*{Acknowledgments}

The generous support provided by the Saudi Center for Theoretical Physics (SCTP)
is highly appreciated by all authors. AJ thanks the Deanship of Scientific Research at King Faisal University for funding this research number  (140232).

\bibliographystyle{plain}

\end{document}